\documentclass[11pt]{article}

\usepackage{amsmath,amsthm,amssymb}
\usepackage{graphicx,psfrag,epsf}
\usepackage{enumerate}
\usepackage{multirow,booktabs}
\usepackage{url} % not crucial - just used below for the URL 
\usepackage{natbib}
\usepackage{xcolor}
\usepackage{colortbl}
\usepackage{makecell}
\usepackage{booktabs}
\usepackage{tabularx}
\usepackage{multibib}
\newcites{refsupp}{Supplementary references}
\usepackage{longtable}
\usepackage{tikz}
\usetikzlibrary{positioning}
\usetikzlibrary{calc}
\usepackage{bm}
\usepackage{hyperref}
\usepackage{geometry}
\usepackage{csquotes}
\usepackage{enumitem}
\usepackage{subcaption}
\usepackage[section]{placeins}
\usepackage{animate}

\makeatletter
\AtBeginDocument{%
	\expandafter\renewcommand\expandafter\subsection\expandafter{%
		\expandafter\@fb@secFB\subsection
	}%
}
\makeatother

\usepackage{etoolbox}
\makeatletter
\patchcmd{\@makecaption}
{\parbox}
{\advance\@tempdima-\fontdimen2} % decrease the width!
{}{}
\makeatother 

\usepackage{geometry}
\usepackage{titling}
\makeatletter
{\begin{titlepage}\def\@thanks{}}%
{\end{titlepage}}
\makeatother

\newtheorem{theorem}{Theorem}

\newtheorem{proposition}[theorem]{Proposition}
\theoremstyle{definition}

\theoremstyle{definition}

\theoremstyle{definition}

\theoremstyle{remark}

\title{ {\huge {Globally aligned Principal Component Analysis for multi-group data}  }\\ \vspace{-1mm}}

\makeatletter
\def\@author{
  \parbox{\textwidth}{\centering
    {\large {Hedayat Fathi}} \\ 
    \small{ Department of Operations and Decision Systems, Universit\'e Laval, Quebec, Canada} \\ 
    \textcolor{blue}{hedayat.fathi.1@ulaval.ca}\\[2mm]
    
    {\large {Marzia A. Cremona}} \\ 
    \small{ Department of Operations and Decision Systems, Universit\'e Laval, Quebec, Canada} \\ \textcolor{blue}{marzia.cremona@fsa.ulaval.ca} \\[2mm]
    
    {\large {Federico Severino}} \\ 
    \small{ Department of Finance, Insurance and Real Estate, Universit\'e Laval, Quebec, Canada} \\ 
    \textcolor{blue}{federico.severino@fsa.ulaval.ca} \\[2mm]
    }
}
\makeatother

\date{}

\begin{document}

\maketitle

\begin{abstract}
We propose a novel principal component analysis (PCA) for multi-group datasets, where the same numerical variables are measured across different groups of observations. Existing approaches either ignore group structure entirely by working with global (pooled) data, focus exclusively on local structure (group-wise PCA), or impose restrictive assumptions of common principal components. Our approach respects the multi-group nature of data while improving global comparability of components. We combine group-specific principal components with global ones through an explicit alignment mechanism based on regularized optimization. We introduce the notion of globally aligned covariance matrix, incorporating weighted contributions from global principal directions in the group-wise covariance matrix. The alignment strength is controlled by regularization parameters that can be tuned to achieve the desired trade-off. Through a comprehensive simulation study, we demonstrate that the proposed aligned PCA achieves a favorable compromise between capturing local variation within groups and maintaining interpretability and stability across groups. Furthermore, in an application to the 2021 Canadian Census socioeconomic data, the proposed aligned PCA yields more comparable and stable region-specific components than pooled or region-wise PCA.
\end{abstract}

\noindent%
{\it Keywords:} dimension reduction, global alignment, multi-group data, principal component analysis, regularization, unsupervised learning

%\newpage

\section{Introduction and literature review}\label{sec:intro}

% -------------------------------------------------------
% General theme and motivation 
% -------------------------------------------------------

Principal component analysis (PCA) is one of the most widely used unsupervised learning techniques to reduce dimensionality and explore data. Since its introduction by \cite{pearson1901} and later formalization by \cite{hotelling1933}, PCA has become a foundational dimension-reduction tool. 
%The method transforms high-dimensional data into a lower-dimensional representation by identifying orthogonal directions that successively maximize variance, making it easier to visualize, interpret, and analyze the data. 
For definitions, fundamental topics, and classic variations, see \cite{jolliffe2002}. For a review of recent developments, see \cite{jolliffe2016}.

The classical PCA assumes that all observations belong to a single homogeneous population. However, in practice, datasets often exhibit a natural grouping structure, with observations collected from distinct populations, regions, time periods, or treatment conditions. We refer to such data as \textit{multi-group data} (or \textit{multi-source data}). The group membership of each observation is encoded by a categorical variable, while the remaining measured variables are continuous.
%Some motivating examples are as follows: \textit{socioeconomic data}, where several variables such as, income, employment, education, etc., are measured across municipalities within provinces or regions. Here, municipalities are naturally grouped by province, but it is important to consider both province-specific patterns and national trends. In \textit{medical research}, clinical studies often include patient groups from multiple medical centers. Each center may have patients with unique characteristics, but it is still important to find patterns that apply across the disease while also considering differences at each site. In \textit{financial data}, asset correlations vary across different market regimes (e.g., normal versus crisis periods), requiring analysis methods that capture regime-specific dynamics while keeping global market interpretability. And finally, in \textit{educational data}, student information is naturally grouped by school or classroom, with within-school variation differing from between-school variation. In these scenarios, applying classical PCA to the entire pooled dataset ignores the group structure and may obscure important group-specific patterns. Conversely, applying PCA separately within each group yields directions that are heterogeneous and completely ignore that each group is part of a larger dataset; they are difficult to compare across groups and fail to leverage information in the global data structure.
Examples include socioeconomic data measured across municipalities within provinces or regions, clinical studies involving several medical centers, financial data observed under different market regimes, and educational data grouped by school or classroom. In such settings, applying PCA to the pooled dataset may obscure important group-specific patterns. Moreover, from a theoretical perspective, even in simple structured models, group dependence is known to introduce systematic bias in principal components \citep{Fan2021}. Applying PCA separately for each group yields directions that are difficult to compare across groups and fail to exploit the fact that all groups belong to a larger common dataset.  
This trade-off has concrete practical consequences. One setting in which it is well documented is the construction of area-based socioeconomic indices, where PCA is among the most widely used methods to derive index weights \citep{mogin2025}. In Canada, for instance, indices such as the Pampalon material and social deprivation index \citep{pampalon2014} and the Canadian Index of Multiple Deprivation \citep[CIMD;][]{statscanada2023cimd} are produced in separate national and regional versions, computed by separate PCAs: the regional versions allow comparison within a region but not across regions, while the national version permits cross-regional ranking but may represent any single region poorly. Such indices are one example of a broader methodological problem -- reconciling group-specific structure with global comparability -- that arises whenever PCA is applied to data with a natural grouping.

% -------------------------------------------------------
%  Main contribution 
% -------------------------------------------------------

In this paper, we propose a simple and computationally efficient approach that bridges the gap between the pooled and the group-wise PCA. Our method, named \textit{globally aligned PCA} (hereafter \textit{aligned PCA}), achieves three objectives simultaneously.
First, it preserves the variation within each group as much as possible by building on the group-wise covariance structure and allowing each group to retain its local variability. Second, it ensures global interpretability. Components are encouraged to align with the global principal components computed from the pooled data, i.e.~the principal angles between the group-specific and global subspaces are encouraged to be small, facilitating cross-group comparison and interpretation.
Finally, it provides interpretable control by introducing a single regularization parameter that controls the strength of alignment, enabling users to explicitly balance group fit against global alignment. 

Our key innovation is the introduction of the \textit{globally aligned covariance} matrix, which adds a low-rank perturbation to each group covariance to increase variation along the selected global directions and thereby encourage alignment with the global subspace. This regularization term has a clear geometric interpretation: it increases the variance along global directions, making them more likely to emerge among the leading principal components of each group.
The proposed method is designed to operate in the intermediate regime between the two extremes of pooled and group-specific PCA. By construction, the group-wise principal components are optimal in the sense that they maximize the explained variance in the group. However, these directions are not designed to be comparable across groups, and cross-group interpretation can be unstable when the leading subspaces vary substantially. Such heterogeneity can arise from several sources: groups may differ in their internal correlation structure or in size, with larger groups dominating the pooled directions while smaller ones are underrepresented. 
In many heterogeneous datasets, there are often alternative directions that explain only slightly less variance of the group-wise components, yet are much more aligned with the pooled (global) principal subspace. Aligned PCA explicitly exploits this trade-off: it tolerates a controlled loss in explained variance of groups in exchange for a substantial gain in similarity to the global directions and stability across groups.
In both the simulation study and the real-data application, appropriate alignment strengths preserve most of the variance of each group while substantially improving global alignment and stability, yielding components that are easier to compare and interpret across groups without collapsing all groups into a single pooled solution.

% -------------------------------------------------------
%  Literature review 
% -------------------------------------------------------

%We now place our contribution within the broader PCA literature.
The literature on PCA extensions is extensive, including robust PCA for data with outliers \citep{candes2011}, kernel PCA for nonlinear structure \citep{scholkopf1998}, sparse PCA to improve interpretability \citep{jolliffe2003, zou2006}, and functional PCA for infinite-dimensional data \citep{dauxois1982, ramsay2005, severino2022}.
Despite the existence of several generalizations to address different shortcomings of classic PCA, methods specifically targeting grouped observation structures remain relatively rare. 
When the covariance matrices $\Sigma_g$ of the groups are assumed to be equal (while the group means may differ), one is in the setting of classical \textit{discriminant analysis}, which distinguishes the groups through their means using this shared covariance \citep[Chapter 9]{jolliffe2002}. 
In this case, the principal components of the groups are identical, being the eigenvectors of the shared covariance matrix, but they may differ from the pooled principal components. 
The method of Common Principal Components (CPC), introduced by \cite{flury1983, flury1988} and simplified by \cite{krzanowski1984}, relaxes this often unrealistic assumption of shared covariance matrix. CPC seeks a single orthogonal matrix $M$ such that $M^\top \Sigma_g M$ is diagonal for all groups. The columns of $M$ are ``common'' principal components, though they may explain different amounts of variance in different groups. Notably, the first CPC component does not necessarily maximize variance in each group. CPC assumes the existence of a global eigenbasis shared by all groups, an assumption that may not hold in heterogeneous settings \citep[see][]{eslami2013general, eslami2014}.  

In the multi-block context, where multiple groups of variables are observed on the same objects, \cite{lock2013joint} proposed Joint and Individual Variation Explained (JIVE), which decomposes the data into joint and individual components. JIVE addresses integration across groups of variables rather than across groups of observations sharing the same variables, as in our proposed method. 
\cite{tenenhaus2014} observed that multi-block methods can also be applied to multi-group data, e.g., by considering the transposition of matrices, and extended regularized generalized canonical correlation analysis (RGCCA) to the multi-group case. The setting most closely connected to ours is their ``Situation 3'', which seeks a single direction from a modified version of the pooled data, and uses it as component of every group. While RGCCA relies on a regularized criterion, it generally lacks a closed-form solution and does not interpolate between the group-wise and global directions via a tunable parameter.
In a very recent contribution, \cite{puchhammer2026} combined sparse and robust PCA for multi-group data, balancing global and local sparsity patterns in the loadings through penalties and a smoothed robust covariance estimator. Although their method also navigates between global and local structure, it does so through shared sparsity patterns and covariance smoothing, whereas our penalty acts directly on the directions, pulling the group-wise components toward the global principal subspace.
In the supervised literature, \cite{chiaromonte2002sufficient} developed sufficient dimension reduction (SDR) for multi-group data. While conceptually related to our approach, SDR differs by incorporating response information.

Regarding the comparison of principal subspaces, \cite{krzanowski1979} developed methods to assess the similarity of principal components across groups using angles between subspaces. Later, \cite{keramidas1987} 
%and \cite{korenius2007} 
proposed to compare the eigenvectors using Euclidean distance or cosine similarity to a reference vector or a ``typical'' component. These methods provide similarity metrics, but do not integrate group-specific and global structure into a single estimation framework.

% -------------------------------------------------------
% Structure of the paper 
% -------------------------------------------------------

The remainder of the paper is organized as follows. Section~\ref{sec:Method} introduces the proposed aligned PCA framework, including the globally aligned covariance construction and the role of the alignment parameters in controlling the local-global trade-off. Section~\ref{sec:simulation} presents a Monte Carlo simulation study to evaluate the method in terms of within-group fit, global alignment, and stability across groups. Section~\ref{sec:application} applies the approach to the 2021 Canadian Census socioeconomic data. Section~\ref{sec:conc} concludes and outlines directions for future work.

\section{Methodology}\label{sec:Method}

This section introduces the globally aligned PCA method. After fixing the notation for group-wise and pooled data, we define the globally aligned covariance matrix, which is a low-rank modification of the group covariance that pulls the group-wise components toward the global principal subspace. We provide an illustrative example and study the limiting behavior of the aligned components as the alignment strength varies.

\subsection{Globally aligned principal components}

Consider a multi-group dataset consisting of $G$ groups, where each group $g \in \{1, \ldots, G\}$ contains $n_g$ observations of $p$ numerical variables, and let $\{ X_{ig} \in \mathbb{R}^p : i =1, \ldots, n_g, g = 1, \ldots, G\}$ denote the full dataset with the total sample size $\sum_{g=1}^{G} n_g = n$.
The group membership of each observation $X_{ig}$ is encoded by a categorical variable with $G$ classes, so that the full data structure consists of $p$ continuous variables together with one categorical variable.
%In applications, we center and standardize each variable using the pooled sample before the analysis. 
We consider the group-wise (for $
g = 1, \ldots, G$) and pooled means
\[
\bar X_g=\frac{1}{n_g}\sum_{i=1}^{n_g} X_{ig}, \quad
\bar X=\frac{1}{n}\sum_{g=1}^{G}\sum_{i=1}^{n_g} X_{ig},
\]
as well as the group-wise (for $
g = 1, \ldots, G$) and pooled covariance matrices
\[
\Sigma_g=\frac{1}{n_g}\sum_{i=1}^{n_g}\left( X_{ig} - 
\bar X_g \right) \left( X_{ig} - \bar X_g \right)^\top,
\quad
\Sigma=\frac{1}{n}\sum_{g=1}^{G}\sum_{i=1}^{n_g} 
\left( X_{ig}-\bar X\right)\left(X_{ig}-\bar X \right)^\top.
\]
Our methodology assumes that the covariance matrices are reliably estimated. When the sample size is small relative to $p$, the sample covariance matrix can be poorly estimated, and a substantial literature addresses this through shrinkage \citep{ledoit2004well} or sparse regularization \citep{bickel2008regularized, cai2011adaptive}. Our penalty is applied on top of, rather than in place of, such estimators. 
Since $\Sigma_g$ and $\Sigma$ are covariance matrices, they are symmetric and positive semidefinite, and therefore, by the spectral theorem, they admit $p$ real non-negative eigenvalues (counted with multiplicity) and an orthonormal eigenbasis of $\mathbb{R}^p$.
Let $(\lambda_{g,k}, v_{g,k})_{k=1}^p$ denote the eigenpairs 
of $\Sigma_g$, ordered so that
\[
\lambda_{g,1} \ge \cdots \ge \lambda_{g,p} \ge 0, \qquad 
\left\|v_{g,k}\right\|_2 = 1, \qquad k = 1, \ldots, p,
\]
and let $(\lambda_m^{\text{global}}, v_m^{\text{global}})_{m=1}^p$ 
be the eigenpairs of $\Sigma$, ordered so that
\[
\lambda_1^{\text{global}}\ge \cdots \ge \lambda_p^{\text{global}}
\ge 0, \qquad
\left\|v_m^{\text{global}}\right\|_2=1, \qquad m = 1, \ldots, p.
\]
The pooled PCA uses $v_1^{\text{global}}, \ldots, v_p^{\text{global}}$ for all groups, while group-wise PCA independently applies $v_{g,1}, \ldots, v_{g,p}$ within 
each group $g$ separately. 
Since $\Sigma$ is computed around the pooled mean, its leading directions summarize the total pooled variation, combining the within-group covariance structure and the between-group differences when the group means are not identical. 
Pooled PCA may miss group-specific patterns, while group-wise PCA captures local structure but ignores that each group belongs to a larger dataset, yielding components that may be dispersed across groups. We propose an intermediate construction that retains each group's local structure while encouraging its leading components to lie closer to the global principal subspace.

Our method is in the spirit of ridge or lasso regressions: as these methods add a penalty to the least squares estimator \citep{hastie2020ridge}, we add a regularization term to the group covariance matrix $\Sigma_g$. 
In particular, we add a low-rank term that targets specific global directions rather than penalizing all directions equally. In the simplest case, when alignment is desired with only the first global principal component, the penalty is proportional to the outer product $v_1^{\text{global}}( v_1^{\text{global}})^\top$, which increases the variance along that single direction.
In the general case, let $r \leq p$ be the number of global components to be considered in the alignment, and let $\rho_1, \ldots, \rho_r \geq 0$ be alignment parameters controlling the influence of each global direction, collected into the vector $\rho = (\rho_1, \ldots, \rho_r)^\top$. We define \textit{the globally aligned covariance matrix} for group $g$ as
\begin{equation}\label{eq:cov_alig}
\Sigma_g^{(\rho)} = \Sigma_g + \sum_{m=1}^{r} \rho_m 
v_m^{\text{global}} \left( v_m^{\text{global}} \right)^\top.
\end{equation}
Equivalently, let $V_r:=[v^{\text{global}}_1 \cdots v^{\text{global}}_r] \in \mathbb{R}^{p\times r}$ be the matrix of the first $r$ global principal directions and $D_\rho := \text{diag}(\rho_1,\dots,\rho_r)$, then
$\Sigma^{(\rho)}_{g} = \Sigma_{g} + V_r D_\rho V_r^\top.$
Note that for any unit vector $x \in \mathbb{R}^p$,
\[
x^\top \Sigma_g^{(\rho)} x = x^\top \Sigma_g x + 
\sum_{m=1}^r \rho_m \cos^2(\theta_{x,m}),
\]
where $\theta_{x,m} \in [0, \pi/2]$ is the angle between $x$ and $v_m^{\text{global}}$. The penalty increases the variance of $x$ proportionally to its squared cosine 
similarity with each global direction. When $x$ is close to $v_m^{\text{global}}$ (small $\theta_{x,m}$), the effect is strongest; when $x$ is orthogonal to all global 
directions, the effect vanishes. In particular, setting $x = v_m^{\text{global}}$ gives 
$\cos^2(\theta_{x,m}) = 1$, so the variance along each global direction $v_m^{\text{global}}$ increases by exactly $\rho_m$. The larger $\rho_m$ is, the stronger the effect, making $v_m^{\text{global}}$ more likely to appear among the leading eigenvectors of $\Sigma_g^{(\rho)}$. This dependence on the size of the penalty is made explicit in Proposition~\ref{prop:eigenvector_convergence}.

From the spectral decomposition point of view, let $\Sigma_g = U_g \Lambda_g U_g^\top$ be the eigendecomposition of the group covariance, where $U_g$ collects the group-wise eigenvectors and $\Lambda_g = \mathrm{diag}(\lambda_{g,1}, 
\ldots, \lambda_{g,p})$. Then
\(
\Sigma_g^{(\rho)} = U_g \Lambda_g U_g^\top + V_r D_\rho V_r^\top,
\)
which is a sum of two spectral structures. Since both are symmetric positive semidefinite, so is $\Sigma_g^{(\rho)}$, and all its eigenvalues remain real and nonnegative. 
Let
\(
(\lambda_{g,k}^{(\rho)},\ v_{g,k}^{(\rho)})_{k=1}^p
\)
denote the eigenpairs of $\Sigma_g^{(\rho)}$ ordered decreasingly. We have that 
\[
\Sigma_g^{(\rho)} v_{g,k}^{(\rho)}=\lambda_{g,k}^{(\rho)} v_{g,k}^{(\rho)},\qquad k=1,\dots,p,
\]
and $\{v_{g,1}^{(\rho)},\dots,v_{g,p}^{(\rho)}\}$ forms an orthonormal basis of $\mathbb{R}^p$ (with eigenvectors defined up to sign, and up to rotations within eigenspaces when eigenvalues have multiplicity larger than 1). We call $v_{g,1}^{(\rho)},\dots,v_{g,p}^{(\rho)}$ the \emph{globally aligned principal components} of the group $g$.

\subsection{A toy example}

To provide a geometric intuition for the alignment mechanism, we consider a two-dimensional toy example with two groups ($G=2$, $p=2$) and set $r=1$. 
%In this case, the alignment parameter reduces to a single scalar $\rho = \rho_1 = \tau w_1$, which we vary directly for simplicity.
The goal is to visualize how the leading group-wise principal components rotate toward the first global principal component as the alignment strength $\rho$ increases, and to highlight the resulting trade-off between the explained variance of each group and the global alignment. 

We generate two Gaussian groups with different means and covariance matrices. Specifically, for each group $g\in\{1,2\}$ we simulate $n_g=200$ observations
$
X_{ig}\sim \mathcal N \left( \mu_g,\Sigma_g \right)
$, $i=1,\dots,n_g$,
with means $\mu_1 = \left( -0.2,\ 0.1 \right)^\top$ and $\mu_2= \left( 2.4,\ 1.2 \right)^\top.$
Each covariance $\Sigma_g$ is constructed by rotating an axis-aligned ellipse:
\[
\Sigma_g = R \left( \theta_g \right) \ \text{diag}\left( s_{g,1},s_{g,2} \right) \ R \left( \theta_g \right)^\top,
\]
where $R(\theta_g)$ is the $2\times 2$ rotation matrix. We set $(\theta_1,s_{1,1},s_{1,2})=(30^\circ,1.6,0.4)$ and $(\theta_2,s_{2,1},s_{2,2})=(-35^\circ,1.0,0.3)$, creating two groups with distinct local directions of largest variability. Explicitly, we have
\[
\Sigma_1=
\begin{pmatrix}
1.30 & 0.52\\
0.52 & 0.70
\end{pmatrix},
\qquad
\Sigma_2=
\begin{pmatrix}
0.77 & -0.33\\
-0.33 & 0.53
\end{pmatrix}.
\]
By construction, $\Sigma_g$ has eigenvalues $s_{g,1}>s_{g,2}$, and an (up to an arbitrary sign) associated orthonormal eigenbasis given by
$
v_{g,1} = \left( \cos\theta_g,\ \sin\theta_g \right)^\top
$ and 
$
v_{g,2} = \left( -\sin\theta_g,\ \cos\theta_g \right)^\top
$.
%The $2\times 2$ rotation matrix (counterclockwise) by an angle $\theta$ is
%\[ R(\theta)= \begin{pmatrix} \cos\theta & -\sin\theta\\ \sin\theta & \cos\theta \end{pmatrix}. \]
%It is orthogonal, i.e., $R(\theta)^\top R(\theta)=I_2$, and $\det(R(\theta))=1$, and it rotates any vector $x \in \mathbb{R}^p$ by $\theta$ θ radians (or degrees, as long as you are consistent).

From the simulated samples, we compute the empirical group-wise covariances $\widehat\Sigma_g$ and the empirical pooled covariance $\widehat\Sigma$. Let $(\widehat\lambda^{\text{global}}_1,\widehat v^{\text{global}}_1)$ be the leading eigenpair of $\widehat\Sigma$, with $\|\widehat v^{\text{global}}_1\|_2=1$. This vector $\widehat v^{\text{global}}_1$ is the reference ``global direction'' shown as dotted line in Figure~\ref{fig:toy}. The group-wise PC1 directions $\widehat v_{g,1}$ (leading eigenvectors of $\widehat\Sigma_g$) are shown as dashed lines. Furthermore, let $\widehat v^{(\rho)}_{g,1}$ be the leading eigenvector of $\widehat\Sigma_g^{(\rho)}$ (the aligned PC1 in group $g$). Increasing $\rho$ progressively favors directions closer to $\widehat v^{\text{global}}_1$, and the leading eigenvector $\widehat v^{(\rho)}_{g,1}$ rotates toward the pooled direction (unless the group structure is already nearly aligned). In Figure~\ref{fig:toy}, $\rho$ increases smoothly from $0$ to $4$, and only the current value of $\rho$ is displayed above the scatter plot.

\begin{figure}[!bt]
\centering
\animategraphics[autoplay,loop,width=1\linewidth]{6}{frame_}{000}{069}
\caption{Toy example illustrating the variance--alignment trade-off for the aligned PC1. Left panel: dotted line is the pooled PC1; dashed lines are the original group-wise PC1s; solid lines are the aligned PC1s as $\rho$ increases. Circular arcs show the angle to the pooled direction. Right panels: PC1 loadings, PVE by each method within each group, and alignment angles as functions of $\rho$.}
\label{fig:toy}
\end{figure}

To make the trade-off explicit, we plot three additional diagnostics that update with $\rho$. The top-right panel contains PC1 loadings. Since $p=2$, the loading vectors correspond to the two coordinates of each unit PC1 direction. This panel shows barplots of the loadings for the global direction $\widehat v^{\text{global}}_1$, the group-wise directions $\widehat v_{g,1}$, and the aligned directions $\widehat v^{(\rho)}_{g,1}$, separately for Groups 1 and 2. This emphasizes how the aligned loadings interpolate between the group-wise and global patterns.
The middle-right panel shows per-group PVE along each direction. 
%For a unit direction $v\in\mathbb R^2$, we define the within-group PVE in group $g$ as$\text{PVE}_g(v) = v^\top \widehat\Sigma_g v ~/\text{tr}(\widehat\Sigma_g).$ 
This panel reports $\text{PVE}_g(\widehat v_{g,1})$ (``Group-wise''), $\text{PVE}_g(\widehat v^{(\rho)}_{g,1})$ (``Aligned''), and $\text{PVE}_g(\widehat v^{\text{global}}_1)$ (``Global''). By construction, $\widehat v_{g,1}$ maximizes $v^\top\widehat\Sigma_g v$ over unit vectors and therefore produces the largest PVE in the group $g$, while the aligned direction typically sacrifices some PVE as it rotates toward the pooled direction. 
Finally, the bottom-right panel visualizes the alignment with the global direction. We quantify this using the classic sign-invariant angle between the aligned PC1 and the global PC1 \citep{krzanowski1979}:
\[
\theta_g(\rho)=\operatorname{arc\cos}\left( \left| \left( \widehat v^{(\rho)}_{g,1} \right)^\top \widehat v^{\text{global}}_1\right|\right)\in[0,\pi/2].
\]
This panel displays $\theta_g(\rho)$ in degrees for each group, and the scatterplot shows a small circular arc at each group mean to visualize this angle geometrically. Decreasing $\theta_g(\rho)$ indicates stronger alignment with the pooled PC1.

Overall, the animation illustrates the central behavior of the aligned method: as $\rho$ increases, the aligned PC1s $\widehat v^{(\rho)}_{g,1}$ become more aligned with the national PC1 by moving toward $\widehat{v}^{\text{global}}_1$ (the angle decreases), while their PVE decreases for each group modestly relative to the group-wise optimum. 
%Although the illustration uses $p=2$, $G=2$, and $r=1$, the same construction applies in higher dimensions and for multiple global directions.

\subsection{Controlling the alignment parameter}\label{subsec:weight}

Throughout this paper, we use two related but distinct notions to describe the alignment of subspaces, defined via principal angles between subspaces \citep{krzanowski1979, yu2015useful}. We say that a group-specific principal subspace is \emph{aligned} with the global subspaces when the principal angles between it and the global principal subspace are small. Separately, we refer to the \emph{stability} of a collection of group-specific subspaces as the smallness of the principal angles between each pair of group subspaces. These two notions are quantified in Section~\ref{subsec:perform} by the indices $A$ and $S$, respectively.
We aims to improve the alignment of each group with the global subspace and, as a by-product, the stability across groups while preserving as much variance of each group as possible. 

As discussed, the matrix $V_r D_\rho V_r^\top$
%$\sum_{m=1}^{r} \rho_m v_m^{\text{global}} (v_m^{\text{global}})^\top$
acts as a ``bias'' toward the global principal subspace. 
The integer $r \le p$ controls the dimension of such a subspace used for alignment. A simple and interpretable choice is $r=1$, which targets only the leading pooled direction $v_1^{\text{global}}$.
This choice is appropriate when the leading pooled component is well separated, e.g., when the eigengap $\lambda_1^{\text{global}}-\lambda_2^{\text{global}}$ is large, or when the primary goal is simply to improve the alignment with respect to the main direction of variability. Another natural data-driven approach is to select $r$ based on the pooled proportion of variance explained (PVE). 
Specifically, for a target level $\gamma\in(0,1)$ (e.g., $\gamma=0.80$), we can choose $r$ as the smallest number of leading global principal components that together explain at least a proportion $\gamma$ of the total variance. We use this criterion in our empirical analysis. 

The parameters $\rho_m$ control the influence of the global directions. To make this influence easier to interpret and adjust, it is convenient to decompose $\rho_m$ into two parts. Specifically, let $w_1,\dots,w_r\ge 0$ be fixed weights that encode the relative importance of the first $r$ global directions, 
%(for example, $w_m$ can be the explained-variance ratio associated with the $m$-th global component),
and for $m=1,\dots,r$ set $\rho_m = \tau~ w_m,$ where $\tau\ge 0$ is a single \emph{alignment-strength} parameter. Geometrically, this is a kind of polar decomposition: the weights $w_m$ fix the direction, while $\tau$ acts as a radius controlling its magnitude. In this parameterization, $\tau=0$ yields the group-wise solution, while increasing $\tau$ progressively encourages the leading group-specific components to lie closer to the global principal subspace. 

A practical advantage of the reparameterization $\rho_m=\tau w_m$ is that it separates the relative emphasis across global directions (through $w_m$) from the overall alignment strength (through $\tau$).
The weights $w_1, \ldots, w_r$ can be chosen in several ways. The simplest choice is uniform weighting, $w_1=\cdots=w_r=1$, so that $V_r D_\rho V_r^\top=\tau V_rV_r^\top$ encourages alignment with the global subspace without favoring any particular global component within it. One can also prioritize well-separated directions by choosing $w_m$ proportional to the eigengap $\lambda_m^{\text{global}}- \lambda_{m+1}^{\text{global}}$, as directions with a large eigengap are more reliably estimated \citep{yu2015useful}. However, it is more natural to let the weights decrease with $m$, so that the most dominant global directions receive stronger alignment. A way to achieve this is to choose $w_m$ proportional to the eigenvalues of the first $r$ global components, that is, $w_m = \lambda_m^{\text{global}}$. For numerical stability and interpretability, one can normalize the weights by setting, e.g., $w_m=\lambda_m^{\text{global}}/ \sum_{\ell=1}^r \lambda_\ell^{\text{global}}$. In our simulation and real-data application, we follow this strategy and tune $\tau$ over a grid to control the variance-alignment trade-off.

Given the weights $w_m$, the scalar $\tau$ can be selected in several ways. We can use \textit{reconstruction-based cross-validation}: we split the data into training and test sets, estimate the aligned components on the training data, 
and choose $\tau$ that minimizes the test reconstruction error (computed with respect to the original variables). Another option is the \textit{stability-based selection}: we choose $\tau$ that optimizes a stability criterion (e.g., the stability index $S$ in Section~\ref{sec:simulation}), favoring components that vary less across groups.
A further option is the \textit{guided selection}, which applies when we have a target trade-off between the explained variance and the global alignment. For a group $g$ and a matrix $V \in \mathbb{R}^{p \times r}$ with orthonormal columns, the proportion of variance of group $g$ explained by the subspace spanned by $V$ is 
$\mathrm{PVE}_g(V) = \mathrm{tr}(V^{\top} \Sigma_g V) / \mathrm{tr}(\Sigma_g)$, 
computed with respect to the original group covariance $\Sigma_g$. In this case, $\tau$ can be set to achieve a prescribed tolerance on the within-group loss while improving alignment to the pooled principal subspace. Given a maximum acceptable relative loss $\epsilon \in [0,1]$ in each group's PVE, we define the optimal alignment strength as 
\begin{equation}\label{eq:tau_star}
\tau^*(\epsilon) = \max \left\{ \tau \geq 0 : \ 
\mathrm{PVE}_g\!\left(V_{g}^{(\rho)}\right) \geq 
(1 - \epsilon) \, \mathrm{PVE}_g\!\left(V_{g}\right) 
\text{ for all } g \right\},
\end{equation}
which selects the largest $\tau$ compatible with the prescribed within-group loss tolerance. In practice, we evaluate $\tau$ over a fine grid and select the largest value that satisfies the PVE-loss constraint in every group. 
A final option is the \textit{domain expertise} for those who are familiar with their data to assess whether weak alignment ($\tau = 0.1$--$1.0$), moderate alignment ($\tau = 1.0$--$3.0$), or strong alignment ($\tau \geq 3$) is appropriate given their goals.

\subsection{Performance criteria}\label{subsec:perform}

To assess the performance of the aligned PCA, and to compare it with the group-wise and global alternatives, we introduce four complementary criteria. The first two measure the explained variance of the groups, and the latter two measure how well aligned and stable the subspaces are with the global structure and with each other. 
%These criteria are used throughout the simulation study and the real-data application.
We compare the following three group-wise directions:
\begin{enumerate}
    \item The \textit{global method} uses the first $r$ global principal components for every group $g = 1, \ldots, G$:
    $V^{\text{glob}}_{g} := V^{\text{global}}_r = 
    [v^{\text{global}}_1, \ldots, v^{\text{global}}_r]$
    \item The \textit{group-wise method} uses $V^{\text{group}}_g$, the leading $r$ eigenvectors of $\Sigma_g$, separately for each group.
    \item The \textit{globally aligned method} uses the aligned covariance $\Sigma_g^{(\rho)}$ from eq.~\eqref{eq:cov_alig}, with $\rho$ set as in Subsection~\ref{subsec:weight}. 
\end{enumerate}

To shorten the notation, we use the superscript $^{(m)}$ to denote each of the global, group-wise, and aligned methods. For each method $m$ and group $g$, let $V^{(m)}_g \in \mathbb{R}^{p \times r}$ be the matrix whose columns are the leading $r$ eigenvectors of the corresponding covariance matrix: the pooled $\Sigma$ for the global method, the group covariance $\Sigma_g$ for the group-wise method, and the aligned covariance $\Sigma^{(\rho)}_g$ for 
the aligned method. %Let $P^{(m)}_g = V^{(m)}_g(V^{(m)}_g)^\top$ be the associated rank-$r$ orthogonal projector. 
We collect these basis matrices into the family $\mathcal{V}^{(m)} = \{V^{(m)}_g: g = 1, \ldots, G\}$, which serves as the argument of the four performance criteria defined below.

We define the \textit{ average within-group variance} as
\begin{equation} \label{eq:sim-W}
    W\left( \mathcal{V}^{(m)} \right) = \frac{1}{G} \sum_{g=1}^{G} 
     \mathrm{tr} \left( \left( V^{(m)}_g \right)^\top 
    \Sigma_g V^{(m)}_g \right),
\end{equation}
and the \textit{average proportion of variance explained} as
\begin{equation} \label{eq:sim-PVE}
    \mathrm{PVE} \left( \mathcal{V}^{(m)}\right) = \frac{1}{G} 
    \sum_{g=1}^{G} 
    \frac{\mathrm{tr}\left( (V^{(m)}_g)^\top 
    \Sigma_g V^{(m)}_g \right)}
    {\mathrm{tr} \left( \Sigma_g \right)},
\end{equation}
which normalizes $W$ by the total variance in each group. For the 
group-wise method, $W(\mathcal{V}^{(\text{group})})$ achieves its 
maximum value, equal to the average sum of the $r$ leading eigenvalues 
of $\Sigma_g$; for any other method, $W(\mathcal{V}^{(m)}) \leq 
W(\mathcal{V}^{(\text{group})})$.

The remaining two criteria are based on the notion of principal angles between subspaces \citep{krzanowski1979, yu2015useful}. For two $r$-dimensional subspaces spanned by matrices $V, V' \in \mathbb{R}^{p \times r}$ with orthonormal columns, the $r$ principal angles $\theta_1, \ldots, \theta_r \in [0, \pi/2]$ are defined via the singular values of $V^\top V'$, and the quantity $\|V^\top V'\|_F^2 = \sum_{j=1}^r \cos^2(\theta_j)$ summarizes them. When the two subspaces coincide, all cosines are equal to 1 and $\|V^\top V'\|_F^2 = r$; when they are orthogonal, all cosines are equal to 0. Both indices below are averages of such quantities, differing only in the reference subspace used for comparison.

To quantify the alignment of group-wise subspaces with the global PCA, we define \textit{the alignment index} as
\begin{equation} \label{eq:sim-A}
A \left( \mathcal{V}^{(m)}\right) = \frac{1}{G} \sum_{g=1}^{G} 
\frac{1}{r} 
    \left\| \left( V^{\text{global}}_r \right)^\top 
    V^{(m)}_g \right\|^2_F
\in [0,1].
\end{equation}
This index is the average, over groups, of the mean squared cosine of the $r$ principal angles between the group subspace and the global $r$-dimensional subspace. If all group subspaces coincide (up to sign) with the global $r$-dimensional subspace, then $A(V^{(m)}) = 1$; if they are all nearly orthogonal to it, $A(V^{(m)})$ is close to $0$.

To quantify the stability of the group subspaces with respect to each 
other, we define the \textit{stability index} as: 
\begin{equation} \label{eq:sim-S}
S\left( \mathcal{V}^{(m)} \right) = \frac{2}{G \left( G-1 \right)}
\sum_{1 \le g < g' \le G} \frac{1}{r}
\left( r - \left\| \left( V^{(m)}_g \right)^\top 
V^{(m)}_{g'} \right\|^2_F \right) \geq 0,
\end{equation}
which is the average, over pairs of groups, of the mean squared sine of the $r$ principal angles between each pair of group subspaces. 
Since each term $(r - \|(V^{(m)}_g)^\top V^{(m)}_{g'}\|_F^2)/r$ is the average squared sine of the principal angles between two subspaces, $S \in [0, 1]$. The index equals zero if and only if all groups share the same $r$-dimensional principal subspace, and equals one if and only if every pair of group subspaces is orthogonal.

Thus, both $A$ and $S$ are averages of similar geometric quantities, but measured against different reference subspaces: $A$ compares each group subspace to the fixed global subspace, while $S$ compares group subspaces to each other. 
Improving $A$ toward $1$ implies that all groups align with the global structure; improving $S$ toward $0$ implies that all groups agree with one another. By construction, perfect global alignment ($A = 1$) implies a perfect stability ($S = 0$), since all group subspaces would coincide with the global subspace and, hence, with each other. The converse does not hold in general: groups can be mutually stable 
($S = 0$) while sharing a subspace that differs from the global one.

\subsection{Limiting behavior}

The following two propositions characterize the behavior of the aligned principal components as the alignment strength $\tau$ varies. Two limits are of interest: $\tau = 0$, which switches off the penalty and recovers the group-wise components, and $\tau \to +\infty$, which makes the penalty dominate the group covariance and gives the leading aligned components toward the global principal directions. However, the latter limit depends on how the alignment weights $w_1,\dots,w_r$ are selected. As $\tau$ grows, the penalty pushes the aligned components toward the global direction carrying the
largest weight, so the order of $w_1,\dots,w_r$ controls which
global direction emerges as the leading aligned component. Ordering the weights as $w_1 > \cdots > w_r$ is precisely what makes the $m$-th aligned component converge to the $m$-th global direction. If, for instance, $w_1 = 0.01$ and $w_2 = 0.99$, the leading aligned component would instead converge to $v_2^{\mathrm{global}}$. When two weights coincide, say $w_1 = w_2$, the individual components
$v_{g,1}^{(\rho)}, v_{g,2}^{(\rho)}$ are no longer separately identified in the limit, since any pair of orthonormal vectors in
$\mathrm{span}\{v_1^{\mathrm{global}}, v_2^{\mathrm{global}}\}$ is a valid choice. The uniform weights $w_1 = \cdots = w_r = 1$ recommended in Subsection~\ref{subsec:weight} fall outside the scope of the following proposition, but can still be used in practice. In that case, the leading aligned components do not individually converge to the global directions
$v_m^{\mathrm{global}}$, but each tends to a unit vector in
$\mathrm{span}\{v_1^{\mathrm{global}}, \dots, v_r^{\mathrm{global}}\}$.
The following proposition formalizes the limiting behavior.
%under the strict ordering $w_1 > \cdots > w_r > 0$. 
The proof of this proposition is provided in the Supplementary Material.

\begin{proposition} \label{prop:eigenvector_convergence}
Assume $\operatorname{tr}(\Sigma_g)>0$, and let $\rho_m = \tau w_m$ for given weights $w_1 > w_2 > \cdots > w_r > 0$. When $\tau = 0$, the aligned principal components coincide with the group-wise principal components. When $\tau \to +\infty$, the leading $r$ eigenvectors of $\Sigma_g^{(\rho)}$ converge, up to a sign, to the global principal directions, that is, $v_{g,m}^{(\rho)} \to v_m^{\mathrm{global}}$ for $m = 1,\dots,r$.
%However, the corresponding eigenvalues diverge, $\lambda_{g,m}^{(\rho)} \to \infty$, so the pooled eigenvectors are recovered in the limit but not the pooled eigenvalues.  
\end{proposition}

While the aligned eigenvectors converge to the global directions, the corresponding aligned eigenvalues behave differently. Since $\Sigma^{(\rho)}_g = \Sigma_g + V_r D_\rho V_r^\top$ adds a positive semidefinite term to $\Sigma_g$, by Weyl's monotonicity theorem \citep[Corollary~III.2.3]{bhatia1997}, no eigenvalue can decrease, and $\lambda^{(\rho)}_{g,k} \ge \lambda_{g,k}$ for all $k=1,\dots,p$.
Moreover, since the columns of $V_r$ are orthonormal, the matrix
$V_r D_\rho V_r^\top$ has the same nonzero eigenvalues as $D_\rho$, namely $\tau w_1, \ldots, \tau w_r$. Applying the same monotonicity argument with $\Sigma_g$ as the positive semidefinite perturbation yields $\lambda^{(\rho)}_{g,m} \ge \tau w_m$ for $m=1,\dots,r$. Therefore, the leading
$r$ aligned eigenvalues diverge as $\tau\to+\infty$.
By contrast, the variance of groups explained by the aligned components remains bounded as $\tau \to +\infty$. In fact, the explained variance must be measured with respect to the \emph{original} group covariance $\Sigma_g$, rather than the modified $\Sigma_g^{(\rho)}$. In other words, the within-group PVE of any direction $v$ is always computed from the original group covariance $\Sigma_g$, independent of the alignment penalty. The following proposition gives explicit bounds and the limiting value of the proportion of variance explained by the aligned principal components. The proof is given in the Supplementary Material.

\begin{proposition}\label{prop:pve_bound}
Assume $\operatorname{tr}(\Sigma_g)>0$, and let $\rho_1,\dots,\rho_r \ge 0$ be arbitrary alignment parameters. For each $m = 1, \ldots, r$, the aligned principal component $v_{g,m}^{(\rho)}$ satisfies
\[
\frac{\lambda_{g,p}}{\mathrm{tr}(\Sigma_g)} \leq \mathrm{PVE}_g(v_{g,m}^{(\rho)})
\leq \frac{\lambda_{g,1}}{\mathrm{tr}(\Sigma_g)}.
\]
Furthermore, if $\rho_m = \tau w_m$ with given weights $w_1 > \cdots > w_r > 0$, then, as $\tau \to +\infty$, 
\[
\mathrm{PVE}_g(v_{g,m}^{(\rho)}) \to 
\frac{\sum_{k=1}^p \lambda_{g,k} 
\cos^2(\alpha_{m,k})}{\mathrm{tr}(\Sigma_g)},
\]
where $\alpha_{m,k}$ is the angle between $v_m^{\text{global}}$ and $v_{g,k}$.
\end{proposition}

The two propositions describe the limiting behavior of the method from complementary angles. Proposition~\ref{prop:eigenvector_convergence} shows that, under appropriate parameterization, the aligned directions converge to the global ones, while Proposition~\ref{prop:pve_bound} shows what this convergence costs in terms of explained variance. 
Because $v_{g,m}^{(\rho)} \to v_m^{\mathrm{global}}$, the proportion of variance of group $g$ explained by the aligned component converges to $\mathrm{PVE}_g(v_m^{\mathrm{global}})$, the proportion of variance that group $g$ exhibits along the global direction. This limit value is fully
determined by the angles between $v_m^{\mathrm{global}}$ and the
group-wise eigenvectors $\{v_{g,k}\}_{k=1}^p$ and coincides with the within-group value $\lambda_{g,m}/\mathrm{tr}(\Sigma_g)$ only when $v_m^{\mathrm{global}} = v_{g,m}$, i.e., when the group is already aligned with the global direction.

\section{Simulation study}\label{sec:simulation}

In this section, we present a simulation study to illustrate how the aligned PCA balances within-group variation and global alignment. We show that, for suitable choices of the alignment parameter~$\tau$, the aligned method preserves most of the within-group variance while producing group-wise directions that are closer to the global principal component and more stable across groups. We further examine how this trade-off evolves as the degree of between-group heterogeneity increases, and we evaluate performance using various criteria.

We consider $G$ groups, where each group $g \in \{1,\dots,G\}$ contains $n_g$ observations with covariance matrix $\Sigma_g$.
We generate data from a multi-group Gaussian model \citep{franks2019} with $r_{\text{true}} \leq p$ common global directions, perturbed independently within each group. We draw $r_{\text{true}}$ orthonormal vectors $u_{0,1}, \ldots, u_{0,r_{\text{true}}} \in \mathbb{R}^p$ uniformly on the unit sphere. These vectors represent the directions of variation shared across groups.
For each group $g \in \{1, \ldots, G\}$ and each direction $j = 1, \ldots, r_{\text{true}}$, we generate an independent perturbation $\eta_{g,j} \sim \mathcal{N}(0, \sigma_\theta^2 I_p)$ and set
\(
u_{g,j} = (u_{0,j} + \eta_{g,j})/\|u_{0,j} + \eta_{g,j}\|_2,
\)
so that $u_{g,j}$ is a unit vector. Then, the resulting vectors within each group are orthogonalized using the Gram-Schmidt procedure, which sequentially projects each vector onto the orthogonal complement of the previously accepted directions and renormalizes, ensuring that $\{ u_{g,1} , \ldots, u_{g,r_{\text{true}}}\}$ forms an orthonormal set. The parameter $\sigma_\theta > 0$ controls how much the group-wise directions deviate from the global direction $u_{0,j}$. Small $\sigma_\theta$ leads to closely aligned groups, while larger $\sigma_\theta$ produces more heterogeneous principal directions. 
Given $u_{g,j}$, we define the population covariance matrix of the group $g$ as
\begin{equation*}
    \Sigma_g = \lambda_{\text{noise}} I_p + \sum_{j=1}^{r_{\text{true}}} 
    \left( \lambda_j - \lambda_{\text{noise}} \right)  u_{g,j} u_{g,j}^\top,
    \label{eq:sim_cov}
\end{equation*}
where $\lambda_1 \geq \cdots \geq \lambda_{r_{\text{true}}} > \lambda_{\text{noise}} > 0$ are the eigenvalues and $\lambda_{\text{noise}}$ is the common noise eigenvalue shared by all remaining $p - r_{\text{true}}$ directions. 

In the simulations, we set $p =10$, $G = 8$, $r_{\text{true}} = 3$ and fix the eigenvalues as $\lambda_1 = 4, \ \lambda_2 = 2, \ \lambda_3 = 1.5, \   \lambda_{\text{noise}} = 1.$ 
%These choices are motivated by the real data application in Section~\ref{sec:application}. 
%where the first four pooled eigenvalues are $3.37, 1.22, 1.12, 0.99$, displaying a dominant first component and a gradual decay.
So, the total variance under our design is
equal to $\lambda_1 + \lambda_2 + \lambda_3 + (p - r_{\text{true}}) \lambda_{\text{noise}} = 14.5$, which yields cumulative proportions of variance explained of approximately $28\%$, $41\%$, and $52\%$, for the first three components, closely mimicking the structure of the census data analyzed in the real-data application (see Section \ref{sec:application}). All three signal eigenvalues exceed $\lambda_{\text{noise}} = 1$, ensuring that each signal direction is genuinely recoverable. 

For each group $g$, and $i=1,\dots,n_g$ we generate independent observations
\(
X_{ig} \sim \mathcal{N} \left( 0,\Sigma_g \right).
\)
Collecting all observations yields a pooled sample of size $n = \sum_{g=1}^G n_g$. From the pooled data, we compute the empirical global covariance matrix $\widehat{\Sigma}$ and its eigenpairs $(\hat{\lambda}^{\text{global}}_m, \hat{v}^{\text{global}}_m)$, $m=1,\dots,p$. 
Similarly, for each group $g$ we compute the empirical group-wise covariance matrix $\widehat{\Sigma}_g$ and its leading eigenpairs. We evaluate each method using the criteria of Section~\ref{subsec:perform}, applied to subspaces $\hat{V}^{(m)}_g$ and the sample covariances $\hat{\Sigma}_g$.
%$(\hat{\lambda}^{\text{group}}_g,\hat{v}^{\text{group}}_g)$.

We perform the simulation with $n_g = 50$ for each group $g \in \{1, \ldots, G\}$ on the heterogeneity grid $\sigma_\theta \in \{0.05, 0.1, 0.3, 0.5\}$, and the alignment strength $\tau \in \{0, 0.5, 1, 2, 3\}$ (with $\tau = 0$ recovering the group-wise method). 
Each configuration is repeated over 200 Monte Carlo replications. In each replication, we draw a new set of global directions $\{u_{0,j}\}$ and group perturbations $\{\eta_{g,j}\}$, we generate new data matrices, and we compute the three group-wise subspaces. Finally, we report Monte Carlo averages of all criteria.

Table~\ref{tab:sim_fixed} summarizes the results for $\sigma_\theta = 0.3$ and $r = 3$. The group-wise PCA attains the highest within-group variance 
%($W = 8.38$, $\mathrm{PVE} = 57.85\%$) 
but produces heterogeneous directions across groups, 
%as reflected by a moderate alignment index $A = 0.54$ and a relatively large stability index $S = 0.64$.
while the global PCA enforces perfect alignment 
%($A = 1$, $S = 0$)
by construction, but at the cost of reduced explained variance of the groups.
%($W = 6.40$, $\mathrm{PVE} = 44.15\%$).
The aligned PCA interpolates between these two extremes as $\tau$ increases: %At $\tau = 0.5$, the method already reduces the stability index substantially ($S = 0.35$) and improves alignment ($A = 0.81$) while sacrificing only a small amount of explained variance of the groups ($W = 7.99$, $\mathrm{PVE} = 55.12\%$). 
 moderate alignment ($\tau = 1$), already improves the stability and alignment indices substantially while sacrificing only a small amount of explained variance, and stronger penalties ($\tau = 2, 3$) push the components close to the global direction 
 %($A = 0.99$ at $\tau = 3$)
 at the cost of a more pronounced reduction in the explained variance of the groups.
 %($\mathrm{PVE} = 47.73\%$).

\begin{table}[!t]
\centering
\small
\begin{tabular}{lcccccc}
\hline
Method & $\tau$  & $W$ & PVE (\%) & $A$ & $S$ \\
\hline
Group-wise PCA & 0   & 8.38 & 57.85 & 0.54 & 0.64 \\
Aligned PCA    & 0.5 & 7.99 & 55.12 & 0.81 & 0.35 \\
Aligned PCA    & 1.0 & 7.52 & 51.83 & 0.93 & 0.16 \\
Aligned PCA    & 2.0 & 7.11 & 49.02 & 0.98 & 0.06 \\
Aligned PCA    & 3.0 & 6.92 & 47.73 & 0.99 & 0.03 \\
Global PCA     & --  & 6.40 & 44.15 & 1.00 & 0.00 \\
\hline
\end{tabular}
\caption{Average over 200 replications for $\sigma_\theta = 0.3$, $r = 3$, $\tau = 0.5, 1, 2, 3$. 
%$W$ is the average variance of groups explained by $r$-dimensional subspaces eq.\eqref{eq:sim-W}; $\mathrm{PVE}$ is the corresponding average of the proportion of variance of groups eq.~\eqref{eq:sim-PVE}; $A$ is the alignment index eq.~\eqref{eq:sim-A}; $S$ is the stability index eq.~\eqref{eq:sim-S}.
}
\label{tab:sim_fixed}
\end{table}

\begin{table}[!b]
\centering
\small
\begin{tabular}{llccccc}
\hline
$\sigma_\theta$ & Method & $W$ & $\mathrm{PVE}$ (\%) 
& $A$ & $S$ \\
\hline
\multirow{3}{*}{0.05}
  & Group-wise PCA      & 8.38 & 57.85 & 0.73 & 0.43 \\
  & Aligned PCA & 7.97 & 54.99 & 0.97 & 0.07 \\
  & Global PCA         & 7.51 & 51.80 & 1.00 & 0.00 \\[4pt]
\multirow{3}{*}{0.10}
  & Group-wise PCA        & 8.38 & 57.85 & 0.70 & 0.47 \\
  & Aligned PCA & 7.91 & 54.60 & 0.96 & 0.08 \\
  & Global PCA         & 7.34 & 50.63 & 1.00 & 0.00 \\[4pt]
\multirow{3}{*}{0.30}
  & Group-wise PCA       & 8.38 & 57.85 & 0.54 & 0.63 \\
  & Aligned PCA & 7.52 & 51.83 & 0.93 & 0.16 \\
  & Global PCA         & 6.40 & 44.15 & 1.00 & 0.00 \\[4pt]
\multirow{3}{*}{0.50}
  & Group-wise PCA       & 8.38 & 57.85 & 0.48 & 0.68 \\
  & Aligned PCA & 7.38 & 50.88 & 0.90 & 0.19 \\
   & Global PCA         & 6.05 & 41.71 & 1.00 & 0.00 \\
\hline
\end{tabular}
\caption{ Average over 200 replications of $W$, PVE (\%), $A$, and $S$ as a function of heterogeneity $\sigma_\theta$, for global PCA, group-wise PCA ($\tau=0$), and aligned PCA ($\tau=1$, $r=3$).}
\label{tab:sim_heterogeneity}
\end{table}

Table~\ref{tab:sim_heterogeneity} reports how the trade-off changes across heterogeneity levels $\sigma_\theta$, at $r = 3$ and $\tau = 1$. Some patterns emerge clearly. First, $W$ and PVE of the group-wise method are constant across all $\sigma_\theta$ values ($W = 8.38$, $\mathrm{PVE} = 57.85\%$). This is expected because the simulation changes the group-specific eigenvectors while keeping the eigenvalue structure fixed across $\sigma_\theta$. So, the optimal explained variance of the groups remains approximately stable. Second, the alignment index $A$ of the group-wise method decreases as $\sigma_\theta$ grows, dropping from $0.73$ at $\sigma_\theta = 0.05$ to $0.48$ at $\sigma_\theta = 0.5$. Similarly, the stability index $S$ worsens from $0.43$ to $0.68$. This confirms that group-wise PCA becomes increasingly unstable across groups as heterogeneity increases. 
Third, the aligned method with $\tau = 1$ consistently improves both $A$ and $S$ relative to the group-wise solution at every $\sigma_\theta$ level. At low heterogeneity ($\sigma_\theta = 0.05$), the aligned method achieves near-perfect alignment ($A = 0.97$, $S = 0.07$) with a small loss in PVE ($54.99\%$ vs $57.85\%$). At high heterogeneity ($\sigma_\theta = 0.5$), the aligned method still yields a substantial improvement in alignment ($A$ increases from $0.48$ to $0.90$), but the cost in PVE is larger ($50.88\%$ vs $57.85\%$), reflecting the stronger tension between local fit and global alignment when groups are very different.

\begin{table}[!b]
\centering
\small
\begin{tabular}{lcccccc}
\hline
 & \multicolumn{2}{c}{$r = 1$} & \multicolumn{2}{c}{$r = 2$} 
 & \multicolumn{2}{c}{$r = 3$} \\
\cmidrule(lr){2-3}\cmidrule(lr){4-5}\cmidrule(lr){6-7}
Method & $A$ & $S$ & $A$ & $S$ & $A$ & $S$ \\
\hline
Group-wise PCA & 0.54 & 0.74 & 0.51 & 0.70 & 0.54 & 0.64 \\
Aligned PCA    & 0.86 & 0.29 & 0.90 & 0.21 & 0.93 & 0.16 \\
Global PCA     & 1.00 & 0.00 & 1.00 & 0.00 & 1.00 & 0.00 \\
\hline
\end{tabular}
\caption{Alignment index $A$ and stability index $S$ for $r \in \{1, 2, 3\}$ at $\sigma_\theta = 0.3$ and $\tau = 1$.}
\label{tab:sim_subspace}
\end{table}

Table~\ref{tab:sim_subspace} gives the comparison of the higher-dimensional subspaces at $\sigma_\theta = 0.3$ and $\tau = 1$. 
%The columns $r = 3$ reproduce the corresponding entries in Table~\ref{tab:sim_fixed}, and are provided for a simpler comparison.
As $r$ increases, the stability index $S$ improves slightly, 
decreasing from $0.74$ at $r = 1$ to $0.64$ at $r = 3$. Higher dimensional subspaces are forced to overlap more, so the normalized average squared sine of their principal angles tend to be smaller. The alignment index $A$ of the group-wise method remains almost constant ($0.51$--$0.54$), indicating that the average principal angle between group and global subspaces does not worsen proportionally with dimension. For the aligned method with $\tau = 1$, both $A$ and $S$ improve relative to the group-wise solution for every $r$. An important result is that the improvement in $A$ is larger for higher dimensions ($A = 0.86$, $A = 0.90$, $A = 0.93$ for $r =1, 2,3$ respectively), showing that the alignment penalty is effective across the full subspace and not only along the leading direction. The stability index $S$ of the aligned method improves with $r$ (from $0.29$ to $0.16$), remaining well below the group-wise values at every level.

Overall, the simulation study shows that the aligned PCA can achieve a favorable compromise between preserving within-group variation and enforcing global alignment of the principal subspaces. For moderate values of~$\tau$, the aligned components remain close to the group-wise eigenvectors while providing more stable and interpretable directions across groups than the group-wise PCA.
The R code implementing the aligned PCA method and reproducing the simulation study is available at 
%a GitHub repository.
\url{https://github.com/HedayatFathi/Globally_Aligned_PCA}.

\section{Real-data application}\label{sec:application}

In this section, we apply the aligned PCA to the real-world socioeconomic data obtained from the 2021 Canadian Census, made publicly available by Statistics Canada.
%Detailed information about municipalities and regions in Canada, along with related visualizations, can be found on the website \cite{censusmapper}. 
The dataset contains several demographic, housing, and income-related variables collected from 5,161 municipalities in Canada. We restrict to municipalities with populations greater than 500 and without missing values, retaining 2,895 municipalities. 
Some variables, such as population size, dwelling counts, and language usage, have skewed distributions and contain outliers. For these variables, we apply the logarithmic transformation $\log (1+x)$.
%using the \texttt{R} base function \texttt{log1p()}. This function computes $\log (1+x)$ and is suitable for variables with small positive values or zeros. 
Since the variables are measured on different scales, we standardize each variable using the pooled sample.
We use the following numerical variables:  \texttt{\detokenize{log_population}}: log-transformed population size;
 \texttt{\detokenize{population_percentage_change}}: population change since the 2016 census;
  %\item \texttt{\detokenize{log_Total_private_dwellings}}: Log-transformed number of private dwellings
 \texttt{\detokenize{log_population_density}}: log-transformed population density (people per  km\textsuperscript{2});
 \texttt{\detokenize{average_age}}: average age;
 \texttt{\detokenize{log_married}}: log-transformed number of married individuals;
\texttt{\detokenize{median_employment_income_in_2020_among_recipients}}: median employment income in 2020;
  %\item \texttt{\detokenize{Median_total_income_of_household_in_2020}}: Median household income in 2020
 \texttt{\detokenize{unemployment_rate}}: percentage of unemployed; and finally 
\texttt{\detokenize{log_official_languages}}: log-transformed count of official language speakers.
  %\item \texttt{\detokenize{log_No_high_school_diploma_or_equivalency_certificate}}: Log-transformed count of individuals without a high school diploma
In Figure~S4 of the Supplementary Material, we present histograms of all numerical variables after preprocessing and transformation. 
We consider two distinct grouping structures: grouping based on geographic regions and grouping based on the presence of migrants. 

\subsection{Grouping by geographic region}

The first analysis is based on regional groupings of Canadian provinces. Canada has 10 provinces and three territories. To allow a meaningful comparison between regions, we reorganize the provinces in five broader geographic groups following the division used by Statistics Canada 
in the construction of the CIMD regional indices 
\citep{statscanada2023cimd}: \textit{Atlantic Canada} (Newfoundland and Labrador, Prince Edward Island, Nova Scotia, and New Brunswick), \textit{Prairie Provinces} (Manitoba, Saskatchewan, and Alberta), \textit{West Coast} (British Columbia), \textit{Quebec}, and \textit{Ontario}. 
%Sometimes Quebec and Ontario are considered as the “Central Canada” region, but we treat them as separate groups to emphasize their large populations and distinct socioeconomic profiles. This distinction enables the analysis to capture their contributions to national patterns more accurately. 
We exclude the three northern territories (Yukon, Northwest Territories, and Nunavut) from the analysis due to their very sparse populations and markedly different socioeconomic contexts. The number of municipalities in each region is summarized in Figure~S3 in the Supplementary Material.

\begin{table}[!t] 
\centering
\small
\begin{tabular}{lcccccc}
\toprule
Method & $r$ & $\tau$ & $W$ & PVE(\%) & $A$ & $S$ \\
\midrule
Group-wise PCA & 3 & 0  & 6.084 & 79.95 & 0.770 & 0.379 \\
Aligned PCA    & 3 & 1   & 5.877 & 77.49 & 0.908 & 0.175 \\
Aligned PCA    & 3 & 2   & 5.515 & 72.92 & 0.977 & 0.045 \\
Aligned PCA    & 3 & 3   & 5.373 & 71.13 & 0.994 & 0.012 \\
Global PCA     & 3 & --  & 5.247 & 69.57 & 1.000 & 0.000 \\
\midrule
Group-wise PCA & 4 & 0  & 6.750 & 88.43 & 0.804 & 0.286 \\
Aligned PCA    & 4 & 1   & 6.546 & 85.93 & 0.984 & 0.032 \\
Aligned PCA    & 4 & 2   & 6.469 & 85.01 & 0.996 & 0.009 \\
Aligned PCA    & 4 & 3   & 6.439 & 84.65 & 0.998 & 0.004 \\
Global PCA     & 4 & --  & 6.368 & 83.78 & 1.000 & 0.000 \\
\bottomrule
\end{tabular}
\caption{Average region-wise variance $W$ and PVE(\%) by the leading $r$-dimensional subspace, alignment index $A$, and stability 
index $S$ for three methods under different choices of $r$ and $\tau$.}
\label{tab:regional_WAS}
\end{table}

We start by computing the standard PCA for the pooled dataset. The first four eigenvalues are 3.37, 1.22, 1.12, 0.99, corresponding to proportions of explained variance of $42.1\%$, $15.2\%$, $14\%$, $12.3\%$ for the first four principal components. Therefore,  the first three components already explain about $71\%$ of the total variance, and the first four explain about $84\%$. Hence, in the following, we focus on $r \in \{3,4\}$ global principal directions and set the alignment weights $\rho_m$ in eq.~\eqref{eq:cov_alig} proportional to the $m$-th eigenvalue of the pooled covariance matrix. Subsequently, for each of the five regions, we perform a separate standard PCA. Then,  for each combination of $r \in \{3,4\}$ and $\tau \in \{1,2,3\}$, with $\rho_m = \tau~ \lambda_{m}^{\text{global}}$, for $m= 1, \ldots, r$, we perform an aligned PCA on the region-specific subset of the data. 
Finally, we compare these solutions using the criteria introduced in Section~\ref{subsec:perform}. $W$ and $\mathrm{PVE}$ 
(eqs.~\ref{eq:sim-W} and~\ref{eq:sim-PVE}) measure the average region-wise variance and proportion of variance explained by the $r$-dimensional aligned subspace, the index $A$ (eq.~\ref{eq:sim-A}) measures the average alignment of each region's $r$-dimensional subspace with the national $r$-dimensional subspace; and the index $S$ (eq.~\ref{eq:sim-S}) measures the pairwise dispersion of region-specific subspaces.

Table~\ref{tab:regional_WAS} illustrates the trade-off between region-wise fit and global alignment. As expected, the group-wise PCA attains the highest average region-wise variance 
%($\mathrm{PVE} = 79.95\%$ at $r=3$ and $88.43\%$ at $r=4$)
but yields relatively low alignment with the national subspace 
%($A = 0.770$ and $A = 0.804$, respectively)
and moderate instability across regions.
%($S = 0.379$ and $S = 0.286$). 
The global PCA enforces perfect alignment 
%($A = 1$, $S = 0$)
by construction but at the cost of reduced region-wise variance. %($\mathrm{PVE} = 69.57\%$ at $r=3$ and $83.78\%$ at $r=4$). 
The aligned PCA interpolates between these two extremes as $\tau$ increases. At $\tau=1$, the alignment index rises substantially from $A=0.770$ to $A=0.908$ and the stability index improves from $S=0.379$ to $S=0.175$, while the region-wise PVE decreases only modestly from $79.95\%$ to $77.49\%$. Stronger penalties ($\tau=2,3$) push the aligned subspaces very close to the global solution %($A=0.994$, $S=0.012$ at $\tau=3$, $r=3$)
at the cost of a more pronounced reduction in region-wise variance. %($\mathrm{PVE}=71.13\%$). 
%Similar patterns hold for $r=4$, where the aligned method achieves near-perfect alignment ($A=0.984$, $S=0.032$) already at $\tau=1$ with only a modest PVE loss ($85.93\%$ vs $88.43\%$). 
The jump in $W$ and PVE between the $r=3$ and $r=4$ blocks is expected: a four-dimensional subspace captures more variance than a three-dimensional one by construction. Overall, these results confirm that the aligned PCA provides an effective and controlled balance between global and regional solutions, and that moderate alignment strengths are sufficient to achieve near-global alignment for this dataset.

\begin{table}[!tb]
\centering
\small
\begin{tabular}{cl rr rrr rrr}
\toprule
& & \multicolumn{2}{c}{Group-wise} 
& \multicolumn{3}{c}{PVE$_a$ (\%)} 
& \multicolumn{3}{c}{$A_a$} \\
\cmidrule(lr){3-4}\cmidrule(lr){5-7}\cmidrule(lr){8-10}
$r$ & Region & PVE$_g$ (\%) & $A_g$ 
& $\tau=1$ & $\tau=2$ & $\tau=3$ 
& $\tau=1$ & $\tau=2$ & $\tau=3$ \\
\midrule
\multirow{5}{*}{3}
 & Atlantic & 79.07 & 0.649 & 77.90 & 61.78 & 55.16 & 0.690 & 0.922 & 0.988 \\
 & Ontario  & 84.94 & 0.928 & 84.14 & 83.79 & 83.62 & 0.992 & 0.997 & 0.999 \\
 & Prairie  & 76.52 & 0.728 & 72.54 & 70.37 & 69.58 & 0.950 & 0.987 & 0.994 \\
 & Quebec   & 82.37 & 0.969 & 81.84 & 81.60 & 81.48 & 0.997 & 0.999 & 1.000 \\
 & West     & 76.88 & 0.576 & 71.01 & 67.08 & 65.81 & 0.909 & 0.979 & 0.991 \\
\midrule
\multirow{5}{*}{4}
 & Atlantic & 87.86 & 0.914 & 86.80 & 86.39 & 86.17 & 0.994 & 0.998 & 0.999 \\
 & Ontario  & 91.13 & 0.845 & 89.80 & 89.45 & 89.29 & 0.994 & 0.998 & 0.999 \\
 & Prairie  & 85.48 & 0.748 & 80.77 & 78.70 & 77.96 & 0.964 & 0.991 & 0.996 \\
 & Quebec   & 88.69 & 0.738 & 87.22 & 87.04 & 86.94 & 0.998 & 0.999 & 1.000 \\
 & West     & 89.02 & 0.777 & 85.05 & 83.47 & 82.88 & 0.971 & 0.992 & 0.996 \\
\bottomrule
\end{tabular}
\caption{Region-specific summaries for the $r$-dimensional subspace. 
PVE$_g$ and $A_g$ are the proportion of variance explained and the alignment index of the group-wise PCA by the r-dimensional subspace; PVE$_a$ and $A_a$ are the corresponding quantities for the aligned PCA at each $\tau$.}
\label{tab:region_grid}
\end{table}

Table~\ref{tab:region_grid} provides a region-wise view. 
%PVE$_g$ and $A_g$ are the proportion of variance explained by the $r$-dimensional subspace, and alignment index of the group-wise PCA, while PVE$_a$ and $A_a$ are the corresponding quantities for the aligned PCA at each $\tau$.
At $r=3$, the regions differ considerably in their alignment with the national subspace. Ontario and Quebec already exhibit high group-wise alignment, %($A_g = 0.928$ and $A_g = 0.969$, respectively),
and their subspaces change very little under alignment, with PVE losses under one percentage point across all $\tau$ values.
%and $A_a$ is essentially 1 at $\tau = 2$. 
In contrast, West and Atlantic have substantially lower alignment, %($A_g = 0.576$ and $A_g = 0.649$), 
and the aligned method yields more pronounced corrections: for the Atlantic, PVE$_a$ decreases from $79.07\%$ to $55.16\%$ at $\tau = 3$, while $A_a$ increases from $0.690$ to $0.988$. Prairie occupies an intermediate position.
%($A_g = 0.728$), with moderate PVE losses and substantial alignment gains. 
At $r = 4$, all five regions achieve high group-wise alignment,
%($A_g \geq 0.738$), 
and even Atlantic, which was the most heterogeneous region at $r=3$, reaches $A_g = 0.914$, suggesting that the fourth global direction captures variation that is relevant to Atlantic Canada but not represented in the first three global components. 
%Across both panels, the aligned method consistently improves $A_a$ at every $\tau$ with only modest PVE losses for the already well-aligned regions, and with larger but controlled corrections for the more heterogeneous ones. 
Taken together, these results confirm that global alignment acts as a gentle regularization in regions already close to the national subspace, and as a stronger corrective in regions whose subspace departs more substantially from the global structure.

\begin{table}[!t]
\centering
\small
\begin{tabular}{c rrrr c rrrr}
\toprule
& \multicolumn{4}{c}{$r = 3$} & & \multicolumn{4}{c}{$r = 4$} \\
\cmidrule(lr){2-5}\cmidrule(lr){7-10}
$\epsilon$ & $\tau^*(\epsilon)$ & PVE(\%) & $A$ & $S$
&  & $\tau^*(\epsilon)$ & PVE(\%) & $A$ & $S$ \\
\midrule
0\%  & 0.00 & 79.95 & 0.770 & 0.379 &   & 0.00 & 88.43 & 0.804 & 0.286 \\
1\%  & 0.30 & 79.58 & 0.827 & 0.300 &   & 0.05 & 88.38 & 0.834 & 0.258 \\
2\%  & 0.40 & 79.35 & 0.842 & 0.277 &   & 0.45 & 87.28 & 0.947 & 0.101 \\
5\%  & 0.70 & 78.44 & 0.881 & 0.217 &   & 0.85 & 86.20 & 0.979 & 0.042 \\
10\% & 1.30 & 76.55 & 0.926 & 0.142 &   & 3.00 & 84.65 & 0.998 & 0.004 \\
\bottomrule
\end{tabular}
\caption{Guided selection of $\tau$: for each $r$ and tolerance $\epsilon$, 
$\tau^*(\epsilon)$ is the largest alignment strength keeping the relative 
$r$-dimensional region-wise PVE loss below $\epsilon$ in every region. 
$A$ and $S$ are the alignment and stability indices at $\tau^*(\epsilon)$. 
The row $\epsilon = 0\%$ is the group-wise PCA reference.}
\label{tab:regional_inverse}
\end{table}
To illustrate the guided selection strategy introduced in eq.~\eqref{eq:tau_star}, we solve the inverse problem of finding the largest alignment strength $\tau^*(\epsilon)$ compatible with a prescribed loss $\epsilon$ in the $r$-dimensional region-wise PVE in every region. Unlike the coarse grid $\tau \in \{1,2,3\}$ used above, the inverse problem and the figures that follow evaluate $\tau$ over a fine grid in the interval $[0, 5]$ with spacing $0.05$. 
Table~\ref{tab:regional_inverse} reports the resulting $\tau^*$ values and the corresponding aggregate metrics for $\epsilon \in \{0\%, 1\%, 2\%, 5\%, 10\%\}$ and $r \in \{3, 4\}$, where the row $\epsilon = 0\%$ corresponds to the group-wise PCA reference. Even a very strict tolerance of $\epsilon = 1\%$ yields a substantial improvement in alignment and stability while preserving most of the region-wise variance. The patterns for r = 4 are similar, reflecting the stronger alignment pressure when acting on four global directions.

\begin{table}[!t]
\centering
\small
\begin{tabular}{clrr rr rr rr}
\toprule
& & & & \multicolumn{2}{c}{$\tau=1$} & \multicolumn{2}{c}{$\tau=2$} & 
\multicolumn{2}{c}{$\tau=3$} \\
\cmidrule(lr){5-6} \cmidrule(lr){7-8} \cmidrule(lr){9-10}
$r$ & Region & $\text{PVE}_g$ & $\alpha_g$ & $\text{PVE}_a$ & $\alpha_a$ & 
$\text{PVE}_a$ & $\alpha_a$ & $\text{PVE}_a$ & $\alpha_a$ \\
\midrule
\multirow{5}{*}{3}
 & Atlantic & 34.52 & 74.73 & 33.60 & 55.12 & 29.97 & 27.76 & 28.57 & 19.19 \\
 & Ontario  & 61.56 & 10.96 & 61.55 & 10.22 & 61.53 &  9.58 & 61.50 &  9.01 \\
 & Prairie  & 38.09 & 25.71 & 38.01 & 22.04 & 37.83 & 19.14 & 37.61 & 16.84 \\
 & Quebec   & 54.45 &  7.39 & 54.44 &  6.49 & 54.42 &  5.78 & 54.39 &  5.21 \\
 & West     & 41.99 & 10.42 & 41.97 &  9.12 & 41.94 &  8.12 & 41.90 &  7.31 \\
\midrule
\multirow{5}{*}{4}
 & Atlantic & 34.52 & 74.73 & 34.22 & 63.59 & 32.11 & 42.01 & 29.73 & 26.06 \\
 & Ontario  & 61.56 & 10.96 & 61.55 & 10.24 & 61.53 &  9.61 & 61.50 &  9.05 \\
 & Prairie  & 38.09 & 25.71 & 38.01 & 22.04 & 37.83 & 19.14 & 37.61 & 16.84 \\
 & Quebec   & 54.45 &  7.39 & 54.44 &  6.49 & 54.42 &  5.79 & 54.39 &  5.23 \\
 & West     & 41.99 & 10.42 & 41.97 &  9.18 & 41.94 &  8.20 & 41.90 &  7.42 \\
\bottomrule
\end{tabular}
\caption{Region-specific summaries for the first principal component. 
For each region, the number of global directions $r$, and alignment strength 
$\tau$: $\text{PVE}_g$ and $\alpha_g$ are the proportion of variance 
explained (in \%) and angle to the national first PC (in degrees) for 
the group-wise PCA; $\text{PVE}_a$ and $\alpha_a$ are the corresponding 
quantities for the aligned PCA.}
\label{tab:region_pc1}
\end{table}

\begin{figure}[!t]
    \centering
    \begin{subfigure}[t]{0.49\textwidth}
        \centering
        \includegraphics[width=\linewidth]{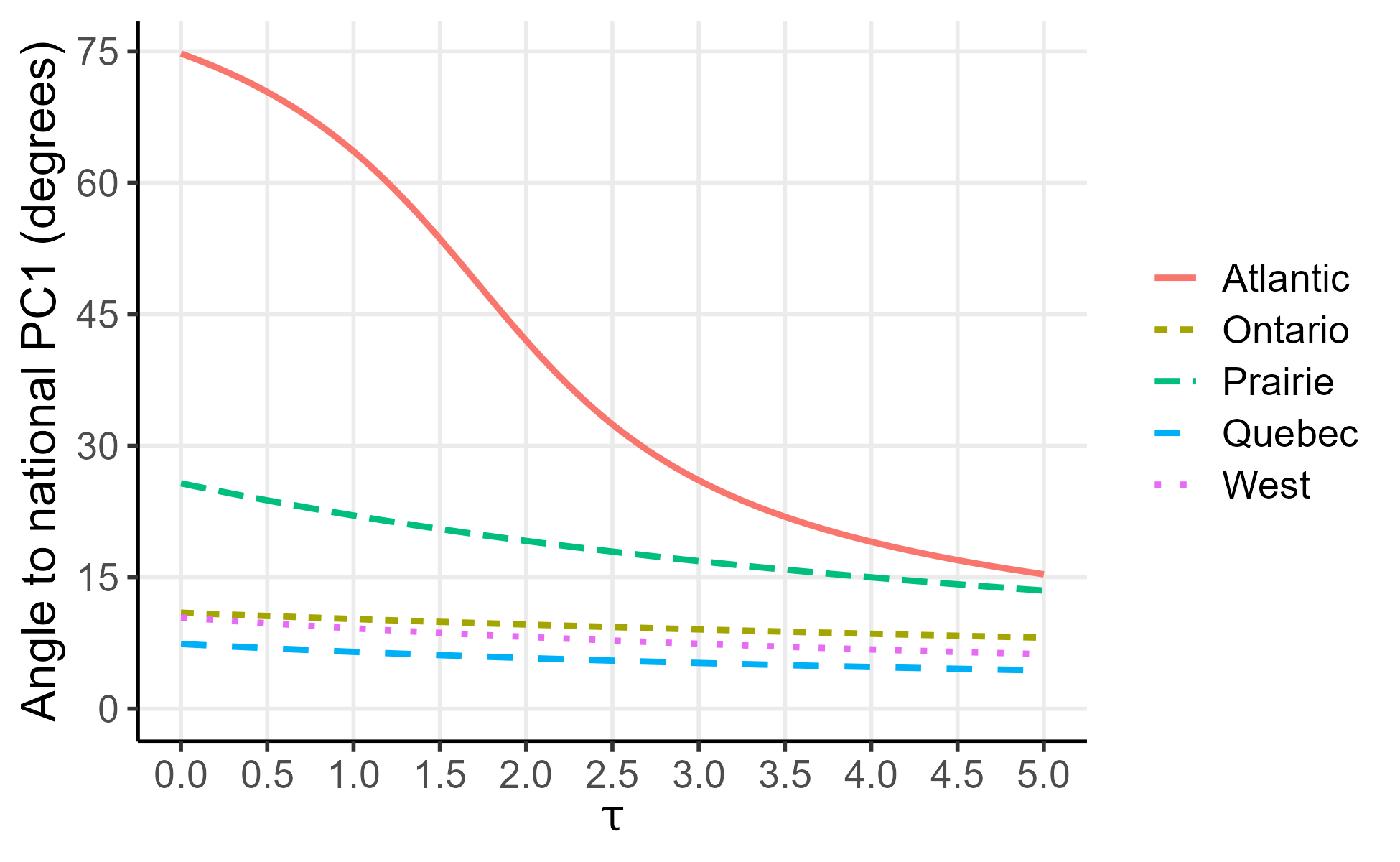}
        \caption{Angle to the national PC1 vs $\tau$}
    \end{subfigure}
    \hfill
    \begin{subfigure}[t]{0.49\textwidth}
        \centering
        \includegraphics[width=\linewidth]{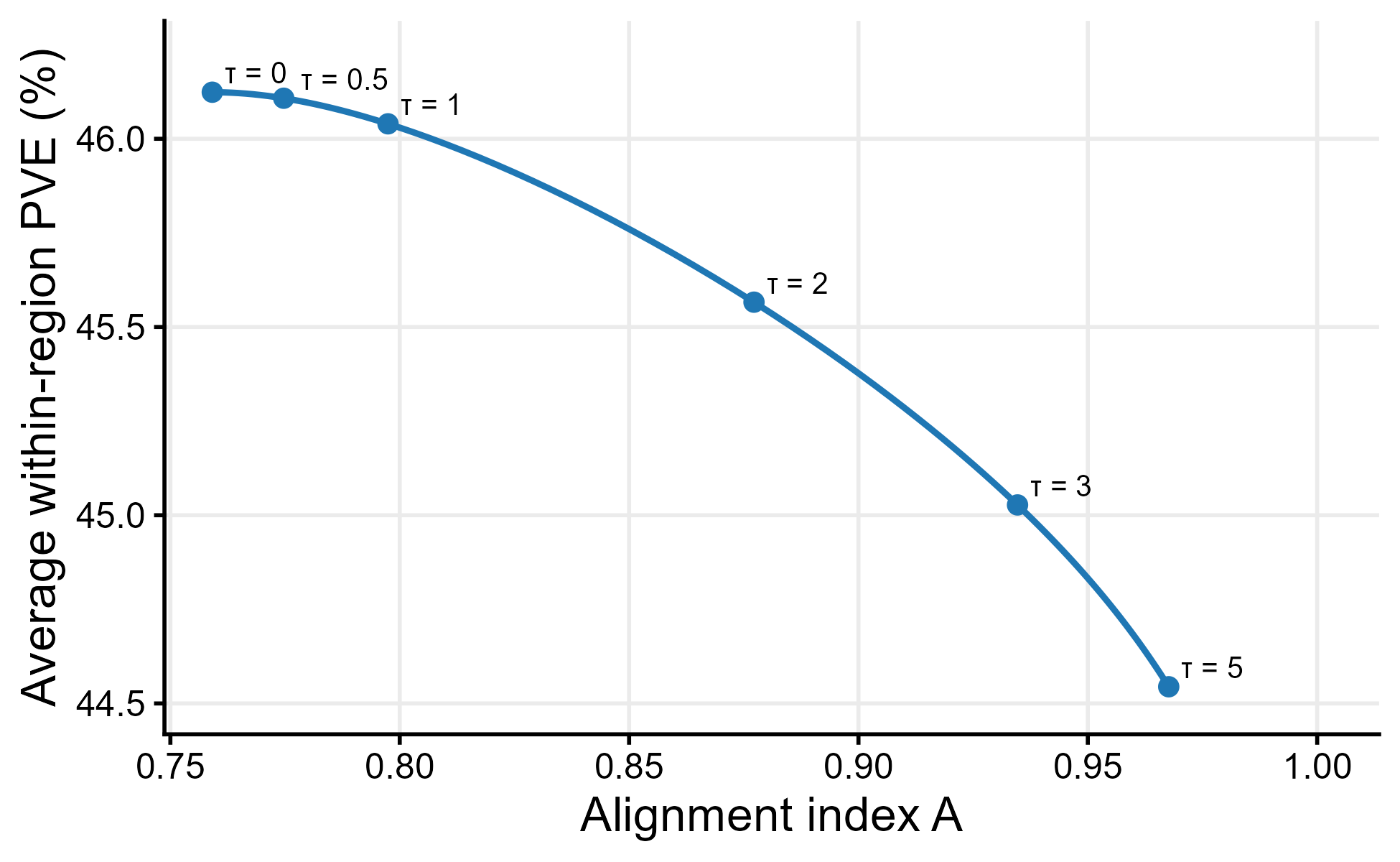}
        \caption{Within-region PVE vs alignment index $A$}
    \end{subfigure}
    \caption{Effect of alignment under the regional grouping 
    (for $r = 4$). (a) Angle between each regional first 
    principal component and the national first principal 
    component as a function of the alignment strength $\tau$. 
    (b) Trade-off between the average within-region PVE and the alignment index $A$ as $\tau$ varies from $0$ to $5$; 
    labeled points mark selected values of $\tau$.}
    \label{fig:regional_trajectory_pareto}
\end{figure}

The remainder of this section focuses on the first principal component, which captures the dominant direction of socioeconomic variation in each region and admits a simple geometric interpretation: the alignment between two directions reduces to a single angle, making the effect of alignment directly visible. 
Table~\ref{tab:region_pc1} provides a region-specific view. 
The group-wise first PC of Atlantic Canada is close to being orthogonal to the national direction ($\alpha_g = 74.73^\circ$), and alignment progressively rotates it toward the national PC: at $r=3$ and $\tau=3$, the angle drops to $19.19^\circ$ at a moderate PVE cost.
%(from $34.52\%$ to $28.57\%$)
%For $r=4$, the correction is more gradual, reflecting the different penalty structure.
The four remaining regions already start close to the national direction 
($\alpha_g \leq 25.71^\circ$), and alignment produces only minor changes. The nearly-identical results for $r = 3$ and $r = 4$ confirm that these regions are robust to the choice of the number of global directions.
Figure ~\ref{fig:regional_trajectory_pareto} summarizes the effect of alignment under the regional grouping. 
The heterogeneous response is clear: the angle for Atlantic Canada decreases sharply toward the national direction, whereas the already-aligned regions start near $10^\circ$ and change only mildly, with the Prairie Provinces showing intermediate behavior. The aggregate trade-off curve is nearly flat for small $\tau$, indicating that substantial gains in alignment can be obtained at almost no cost in within-region variance, and becomes steeper only for larger $\tau$. A region-by-region version is provided in Figure S5 in the Supplementary Material.

\begin{figure}[!tb]
    \centering
    \includegraphics[width= 1 \linewidth]{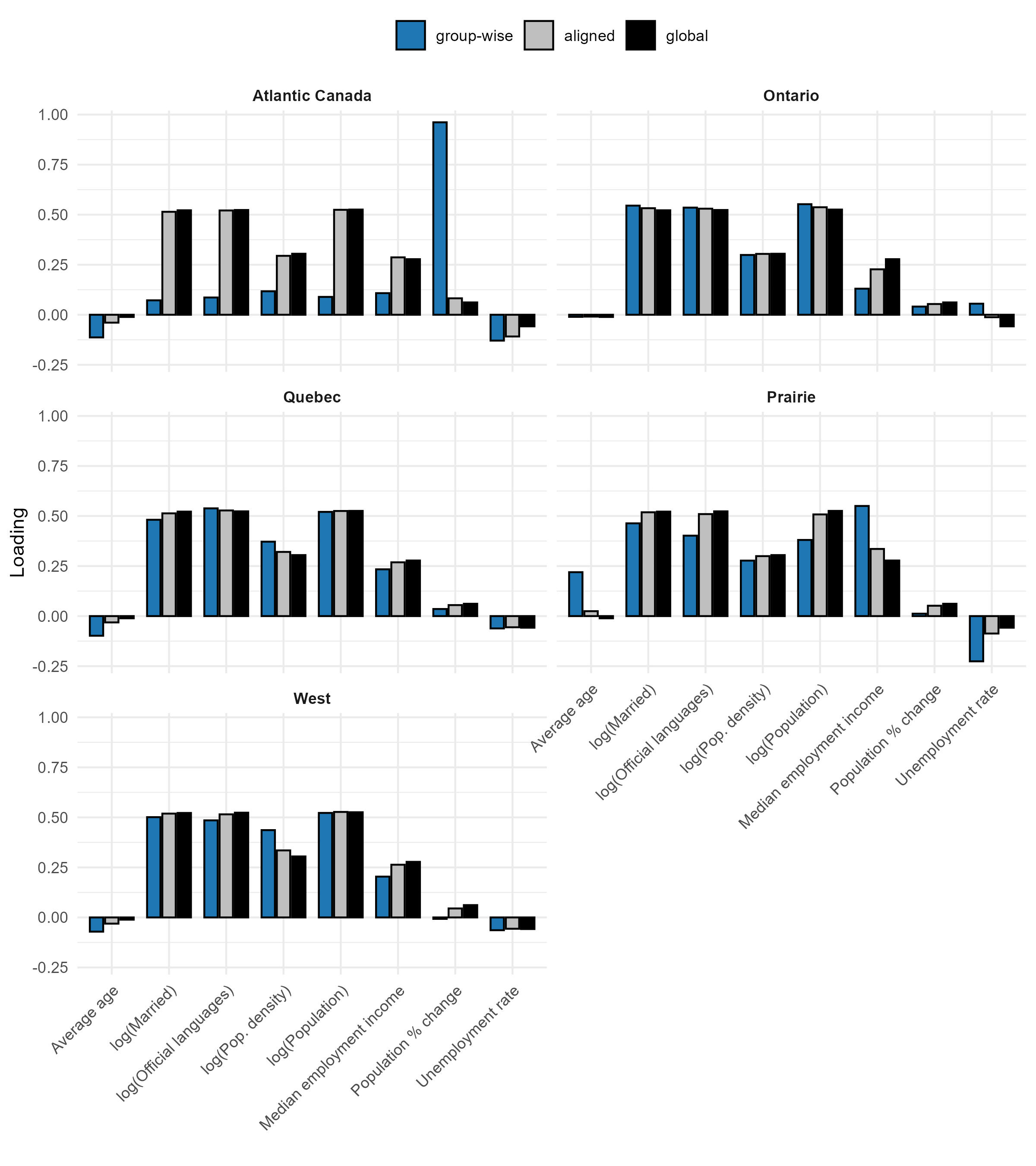}
    \caption{Comparison of the loadings of the first principal component in the five regions for the group-wise PCA, the aligned PCA with $r=4$ and $\tau=3$, and the global PCA.}
    \label{fig:PC1_loadings_comparison}
\end{figure}

To visualize the effect of global alignment on the interpretation of the first component, we also provide plots of loadings for group-wise and aligned PCA, showing that, after alignment, the loadings of each group become closer to the global trend of the overall data. We only plot $r=4$ and $\tau =3$. Among the candidate values of $\tau$, the choice $\tau =3$ corresponds to a relatively strong alignment penalty which, according to Table~\ref{tab:regional_WAS}, substantially increases the alignment and stability indices while still preserving most of the within-region variance. Therefore, Figure~\ref{fig:PC1_loadings_comparison} reports, for this configuration, the loadings of the first principal component for each region, comparing the global PCA, the group-wise PCA, and the aligned PCA.

The loading plots in Figure~\ref{fig:PC1_loadings_comparison} confirm the numerical findings of Tables~\ref{tab:regional_WAS} and~\ref{tab:region_grid}. We see that, in all regions, the loading bars after alignment are closer to the global loadings. In Ontario and Quebec, standard principal components are already well-aligned with the first global principal component. 
%This means that these regions closely follow the national structure and have a strong contribution to the overall variance. 
In contrast, in the Atlantic region, we observe a notable shift in loadings after alignment. For example, the bar representing the median employment income becomes much closer to the global pattern. In the Prairie Provinces and the West Coast regions, the aligned loadings adjust the group-wise principal components only slightly. This suggests that the original structure already reflected the global trend to some extent. Overall, the plot confirms that our method encourages a more aligned representation of regional variation while still preserving meaningful local differences.

\subsection{Grouping by migrant presence}

The second analysis explores a binary grouping based on the presence of external migrants, where municipalities are assigned to two categories: \emph{Migrants} (non-zero external migrants) and \emph{No Migrants} (zero external migrants). The two groups are well represented in the data: the “No Migrants” category contains 1,676 municipalities, and the “Migrants” category 1,219 municipalities. We repeat the global, group-wise, and aligned PCA comparison for this binary grouping, considering $r \in \{3,4\}$ global directions. Since the migrant grouping is binary and less heterogeneous, it aligns rapidly with the national subspace, so we use a finer grid $\tau \in \{0.2, 0.5, 1\}$. As in the regional analysis, the group-wise PCA achieves the largest average group-wise variance but exhibits relatively poor alignment with the national subspace ($A = 0.667$ at $r = 3$ and $A = 0.820$ at $r =4$) and high instability between the two groups. As the alignment strength $\tau$ increases, the aligned PCA trades a modest amount of explained variance for a substantially improved global structure, reaching near-perfect alignment at $\tau =1$ ($A = 0.992$ at $r= 3$ and $A =0.994$ at $r=4$). At the group level, the \emph{Migrants} group already follows the national pattern and requires only minimal adjustment, whereas the more atypical \emph{No Migrants} group is progressively rotated toward the national direction, so that global alignment acts as a gentle adjustment for the former and a stronger corrective for the latter. Overall, the migrant-based grouping confirms that the aligned PCA provides a controlled compromise between local fit and global comparability, even when only two groups are considered. 
The full analysis of this binary grouping, including the detailed numerical summaries, the guided selection of $\tau$, and the loading plots of the first principal component, is provided in Section~S2 of the Supplementary Material.

\section{Conclusion and future work}\label{sec:conc}

We have defined a globally aligned principal component analysis method for multi-group data that extends classical PCA by incorporating explicit alignment with the global structure through a regularization mechanism. The approach is theoretically motivated by regularization methods, computationally tractable via standard eigendecomposition, and practically interpretable through transparent trade-off parameters. The method addresses a fundamental gap in the PCA literature by providing a unified framework that preserves within-group variance and heterogeneity, enforces global interpretability and cross-group comparability, and allows the user to explicitly control the balance through tuning parameters. Simulation studies demonstrate that moderate alignment strengths can achieve favorable compromises, preserving a very large proportion of within-group variance while substantially improving stability and global alignment. 

Future work could extend this framework in several directions. Nonlinear extensions are a future direction. For example, in kernel PCA, one can define globally aligned group covariance operators by adding a low-rank penalty along the leading pooled eigenfunctions in the associated reproducing kernel Hilbert space. Generalizing this technique to functional data is another possible line of research. Moreover, in many applications, the results of PCA are used as inputs to supervised machine learning methods, such as regression analysis. A study of the impact of alignment on subsequent supervised methods can reveal its power. Finally, for classification, the aligned principal components provide a natural dimension-reduction step before linear discriminant analysis, potentially improving robustness by producing features that are more comparable across groups.

\section*{Acknowledgments}
We thank 
Federico Camerlenghi, Tao Chen, Francesca Chiaromonte, Aida Eslami, Patrick Groenen, Ana Maria Kenney, Michael Morin, Mikhail Nediak,  Ndeye Niang,  Laura M. Sangalli, and Piercesare Secchi for useful comments and
the audiences at the Statistical Society of Canada Annual Meeting (SSC 2026), the Canadian Operational Research Society Annual Conference (CORS 2026), the International Federation of Classification Societies Conference (IFCS 2026), and the Data Science, Statistics \& Visualization Conference (DSSV 2026) for helpful discussions.

\section*{Funding}
H.F. was supported by the Faculty of Business Administration, Université Laval, and by the Interuniversity Research Center on Enterprise Network, Logistics and Transportation (CIRRELT). 
M.A.C. is the chairholder of the Chair in Statistical Learning and was funded by the Natural Sciences and Engineering Research Council of Canada (NSERC, grant RGPIN-2020-05657), the Fonds de recherche du Québec Santé (FRQS, grant 2023-2024-JC-339901) and by the Faculty of Business Administration, Université Laval.
F.S. was funded by the Natural Sciences and Engineering Research Council of Canada (NSERC, grant RGPIN-2025-05058) and by the Faculty of Business Administration, Université Laval.

%%%%%%%%%%%%%%%%%%%%%%%%%%%%%%%%%%%%%% REFERENCES
%\clearpage
%\nocite*
%\bibliographystyle{plain}
\bibliographystyle{Chicago}
\small{\bibliography{references}}

\end{document}